\documentclass[reprint,twocolumn,secnumarabic,amssymb, nobibnotes, aps, pra,superscriptaddress]{revtex4-1}

\usepackage{extarrows}
\usepackage[export]{adjustbox}
\usepackage{amsmath}    % need for subequations
\usepackage{graphicx}    % need for figures
\usepackage{verbatim}    % useful for program listings
\usepackage{color}           % use if color is used in text
\usepackage{subfigure}   % use for side-by-side figures
\usepackage{braket}
\usepackage{hyperref}    % use for hypertext links, including those to external documents 
\usepackage{verbatim}  % and URLs
\usepackage{here}
\usepackage{mathrsfs}
\usepackage[english]{babel}
\usepackage[utf8x]{inputenc}
\usepackage[T1]{fontenc}
\usepackage{comment}

\graphicspath{{Figures/}}
\usepackage{physics}
\usepackage{siunitx}
\usepackage{bm}
\usepackage[colorinlistoftodos]{todonotes}

\newcommand{\bR}{\mathbf{R}}
\newcommand{\bS}{\mathbf{S}}
\newcommand{\tS}{\tilde{S}}
\newcommand{\bs}{\mathbf{s}}
\newcommand{\bN}{\mathbf{N}}
\newcommand{\bM}{\mathbf{M}}
\newcommand{\bk}{\mathbf{k}}
\newcommand{\bq}{\mathbf{q}}

\newcommand{\bh}{\bm{h}}
\newcommand{\x}{\operatorname{x}}

\definecolor{blue(pigment)}{rgb}{0.2, 0.2, 0.6}
\definecolor{blue-violet}{rgb}{0.54, 0.17, 0.89}

\begin{document}
%\title{Antiferromagnetic magnons in external magnetic field: dynamical mean-field study}
\title{Antiferromagnetic magnons and local anisotropy: dynamical mean-field study}

\author{A. Niyazi}
\affiliation{Institute of Solid State Physics, TU Wien, 1040 Vienna, Austria}
\author{D. Geffroy}
\affiliation{Department of Condensed Matter Physics, Faculty of
  Science, Masaryk University, Kotl\'a\v{r}sk\'a 2, 611 37 Brno,
  Czechia}
\affiliation{Institute of Solid State Physics, TU Wien, 1040 Vienna, Austria}
\author{J. Kune\v{s}}
\affiliation{Institute of Solid State Physics, TU Wien, 1040 Vienna, Austria}
%\affiliation{Institute of Physics, Czech Academy of Sciences, Na Slovance 2, 182 21 Praha 8, Czechia}
\date{\today}

\begin{abstract}
 We present a dynamical mean-field study of antiferromagnetic magnons
 in one-, two- and three-orbital Hubbard model of square and bcc cubic
 lattice at intermediate coupling strength. We investigate the effect
 of anisotropy introduced by an external magnetic field or single-ion
 anisotropy. 
 For the latter we tune continuously between the easy-axis and easy-plane models. We also analyze
 a model with spin-orbit coupling in cubic site-symmetry setting.
 The ordered states as well as the magnetic excitations are sensitive to even a small breaking
of $SU(2)$ symmetry of the model and follow the expectations of
spin-wave theory as well as general symmetry considerations.
\end{abstract}

\maketitle

\section{Introduction}
The scalability to multi-orbital systems made the dynamical mean-field
theory (DMFT)~\cite{Metzner1989,Georges1992,Georges1996} a widely
used tool for the investigation of electronic correlations on material
specific level ~\cite{Kotliar2006,Held2007}. The primary niche of DMFT
are the one-particle correlation functions (1PCFs) such as the generalized
band structures. Nevertheless, DMFT allows the calculation of higher order
correlation functions as well. In particular, two-particle
correlation functions (2PCFs) play a crucial role in the description of
continuous phase transitions and, in the form of transport
coefficients, dynamical structure factors or susceptibilities are
directly related to numerous experiments. The solution of 
the Bethe-Salpeter equation (BSE) makes the calculation of 2PCFs
numerically very demanding. The DMFT calculations of 2PCFs have so far
been limited to simple models and high
symmetry~\cite{Jarrell1992,Boehnke2011,Hafermann2014,vanLoon2014,Stepanov2018},
or have involved substantial approximations~\cite{Park2011}. Numerical
and analytical developments in representation of
2PCFs~\cite{shinaoka2020} and solution of
BSE~\cite{Wallerberger2020,Otsuki2019,Krien2019,Katanin2020} make
calculations for realistic models with three or more orbitals and
several atoms in the unit cell feasible. 

The collective modes described by 2PCFs play a particularly important role in ordered phases
with spontaneously broken symmetry. Presence or absence of a gap in 2PCF spectrum related
to the type, discrete vs continuous, of broken symmetry is decisive for 
finite-temperature stability of the ordered state in two dimensions~\cite{Mermin1966}.
Therefore it is important to understand not only the analytic
properties of a given theory, for DMFT see
Refs.~\onlinecite{vanLoon2015,Krien2017,krien2018}, but also the 
properties of the actual numerical implementation. Recently, some of us
demonstrated~\cite{geffroy2019, niyazi_dynamical_2020} that the 
dynamical susceptibilities obtained using DMFT respect the Goldstone
theorem~\cite{Goldstone1961, Nambu1961, Goldstone1961}
%analytically in Refs.~\onlinecite{Krien2017,krien2018}
%and numerically 
in the case of $U(1)\times U(1)$ symmetry breaking in spinful excitonic condensate
~\cite{kunes2014c,kunes2014b,nasu2016}.

The most common and largely studied long-range order in strongly
correlated materials is the antiferromagnetic (AFM)
one~\cite{Anderson1952}. The AFM Heisenberg
model~\cite{Manoudakis1991}, which describes fluctuating spins in Mott
insulators, was studied in detail with
analytical~\cite{Chakravarty1989} as well as with numerically exact
methods~\cite{Sandvik1997}. Investigation of antiferromagnetism in
the fermionic Hubbard model, which allows for the description of doped Mott
insulators or AFM metals, relies on weak-coupling methods such as
the random phase approximation~\cite{Rowe2012,DelRe2021}, two-particle
self-consistent approximation~\cite{Vilk1994} or fluctuation-exchange
approximation~\cite{Bickers1989}, numerical simulations on finite
systems~\cite{Luschner2009,Varney2009} or local approximations to
interaction vertices such as DMFT and its
cluster~\cite{Kent2005,Fuchs2011} or diagrammatic
extensions~\cite{Hirschmeier2015,Rohringer2018}.

In this Article, we use DMFT and 2PCFs to study AFM ordering in
the half-filled Hubbard model with one, two and three orbitals. 
While the magnetic phase diagram of the single-orbital Hubbard model
in three dimensions (3D) has been investigated with a number of
methods~\cite{Kent2005,Fuchs2011,Rohringer2011} including
DMFT~\cite{Ulmke1995,Kent2005,Rohringer2011}, we calculate the magnon
spectra and we introduce various symmetry breaking terms such as
external field or single-ion anisotropy, in order to analyze their
effect on magnon dispersions. The calculations are performed for 2D
and 3D lattices. Our goal is to demonstrate the utility of the present
approach for the study of the magnetically ordered phases of
multi-orbital models. In the 2D case, where spontaneous symmetry
breaking is forbidden in the case of continuous
symmetry~\cite{Mermin1966} but allowed in case of discrete symmetry,
we investigate a model, which can be continuously tuned between these
cases.

\section{Computational Method}
The studied Hamiltonian consists of the inter-site hopping, diagonal in
orbital and spin indices, and the on-site part $H_i$ consisting of the
electron-electron interaction, external Zeeman field and spin-orbit
coupling (the specific form will be given later for each studied
case)
\begin{equation}
\label{eq:model}
  H = t\sum_{<ij>,\sigma} \sum_{l=1}^N c_{il\sigma}^{\dagger}
  c_{jl\sigma}^{\phantom\dagger}  +  \sum_{i}H_i.
\end{equation}
Here, $c^{\dag}_{il\sigma}$ and $c_{il\sigma}^{\phantom\dagger}$ are
the fermionic creation  and annihilation operators for electrons with
spin $\sigma$ in orbital $l$ at site $i$ of a square or bcc cubic
lattice. The number of orbitals $N$ ranges from 1 to 3 in the studied
models. For later use we define the occupation number operator
$n_{il\sigma} \equiv
c^{\dag}_{il\sigma}c_{il\sigma}^{\phantom\dagger}$ and the local spin
operators $S^\alpha_{i}\equiv\sum\limits_{l\nu\nu'}
\sigma^{\alpha}_{\nu\nu'}c^{\dag}_{il\nu}c_{il\nu'}^{\phantom\dagger}$,
where $\sigma^\alpha$ are the Pauli matrices. For the sake of
consistency with previous work~\cite{Kunes2011} we choose $t=1/8$
and use it for all studied cases.

The calculations follow the standard DMFT procedure. The lattice model is
mapped onto an auxiliary Anderson impurity model with
self-consistently determined parameters
~\cite{Georges1992,Jarrell1992}, for which the 1PCFs are evaluated
using the ALPS implementation~\cite{Bauer2011,
  Shinaoka2016,Gaenko2017} of the strong-coupling continuous-time
quantum Monte-Carlo (CT-QMC) algorithm~\cite{Werner2006a}. The model
hosts two competing phases: the normal paramagnetic one and the AFM
phase with a staggered spin configuration characterized by the  N\'eel
vector $\bN=\tfrac{1}{2}\langle \bS_A-\bS_B\rangle$. We focus on the 
dynamical spin susceptibility
%\begin{equation}
%\label{eq:operators}
%\begin{split}
%\phi_i^\gamma  & =   %R_i^\gamma+iI_i^\gamma=
%  \sum_{\alpha\beta} %\sigma^\gamma_{\alpha\beta}
%  a_{i \alpha}^{\dagger}
%    b_{i\beta}^{\phantom\dagger}\\
%  O_{i} &= \sum_{\sigma} %(n^{a}_{i\sigma} - n^{b}_{i\sigma}) \\
%//  S_{i}^z&=\sum_{c=a,b}(n^{c}_{i\upar%r//ow} - n^{c}_{i\downarrow}).
%//  \end{split}
%//  \end{equation}
% is characterized by a spontaneous coherence between the $\textit{a-}$
% and $\textit{b}$ -electrons in (1). The correponding order parameter
% is a complex vector with coordinates 
%\begin{equation}
%     \Braket{ \textit{a}_{i \sigma}^{\dagger} %    \textit{b}_{i \bar{\sigma}} }
%\end{equation} % \begin{equation}
%  \phi^\gamma \equiv  \sum_{\alpha\beta} \sigma^\gamma_{\alpha\beta}
%  \expval{a_{\bm{i} \alpha}^{\dagger}
%    b_{\bm{i}\beta}^{\phantom\dagger}},
% \end{equation}
% As shown in the phase diagram transition between
% different phases are realised by varying parameters. Fig.(1) shows the different
% phases at fixed crystal field, which is set to $\Delta = 3.4$ (eV) in
% our calculation, transition is achieved either varying band asymmetry
% $\xi$ or changing temperature T. Fig.(2) is the counterpart at fixed
% band asymmetry, transition is realised by changing crystal field
% $\Delta$ or temperature T.\\
$\chi^{\alpha\alpha}(\bq,\omega)$, which is
obtained by analytic continuation from its Matsubara representation
\begin{equation}
\label{eq:susc}
\frac{1}{N}\expval{\tS^\alpha_{-\bq}\tS^\alpha_{\bq}}_{\omega_n}=
%\chi^{\alpha\alpha}(\bq,i\omega_n)=
\frac{1}{N}\int_0^{\beta}\!\!\!\!\!\mathrm{d}\tau 
e^{i\omega_n\tau} 
\expval{\tS^\alpha_{-\bq}(\tau)\tS^\alpha_{\bq}(0)}.
%-| \expval{S^\alpha_\bq} |^2.
\end{equation}
Here $\expval{X}=\frac{1}{Z}\Tr Xe^{-\beta H}$ denotes the
thermal average, $N$ is the number of lattice sites
and $\tilde{X}=X-\expval{X}$.
We also evaluate the 1P observables such as
the static magnetization $\expval{\bS_\bR}$ and the 1P spectral functions.
The analytic continuation employs the maximum entropy method~\cite{Gubernatis1991},
for details of the present implementation see Refs.~\onlinecite{geffroy2019,Kaufmann2021b}.

The calculations are performed in a two-site unit
cell where the sites are labeled by the sublattice index ${s=A,B}$. The
reciprocal space operators in Eq.~\ref{eq:susc} are then given by
${\tS^\alpha_{\bq}=%\frac{1}{\sqrt{N}}
\sum_{\bR}
e^{-i\bq\cdot\bR}\left(\tS^\alpha_{\bR,A}
+e^{-i\bq\cdot\bs}\tS^\alpha_{\bR,B}
\right)}$, where the sublattice vector $\bs$ assumes the value
$(\tfrac{1}{2},\tfrac{1}{2})$ or
$(\tfrac{1}{2},\tfrac{1}{2},\tfrac{1}{2})$ for the 2D and 3D models,
respectively. The correlation function in Eq.~\ref{eq:susc} is then
obtained as a linear combination of the sublattice contributions
\begin{equation*}
\begin{split}
    \expval{\tS_{-\bq}\tS_\bq}_\omega =  \expval{\tS_{-\bq,A}\tS_{\bq,A}}_\omega +
    \expval{\tS_{\bq,B} \tS_{-\bq,B}}_\omega\\
    +e^{i\bq\cdot\bs} \expval{\tS_{-\bq,B} \tS_{\bq,A}}_\omega
    +e^{-i\bq\cdot\bs} \expval{\tS_{-\bq,A} \tS_{\bq,B}}_\omega.
\end{split}
\end{equation*}
Each term is a contraction of the generalized susceptibility 
$\chi_{ijs,kls'}(\bq,\omega)$
over the spin-orbital indices
\begin{equation*}
\begin{split}
  \frac{1}{N}  \expval{\tS^\alpha_{-\bq,s}\tS^\alpha_{\bq,s'}}_\omega=
    M^\alpha_{\underline{ij}} M^\alpha_{\underline{kl}}
    \chi_{\underline{ij}s,\underline{kl}s'}(\bq,\omega).
\end{split}
\end{equation*}
with matrix elements $M^\alpha_{ij}$ following from the definition of
the spin operators above. To simplify the notation we use underline to
indicate summation over the corresponding indices~\footnote{In case of
summation over the reciprocal lattice vectors this amounts to
$\frac{1}{N}\sum_\bk$, in case of summation over Matsubara frequencies
to $T\sum_\nu$ where $T$ is the temperature.}. 

The calculation of the generalized susceptibility within
DMFT~\cite{Jarrell1992, Georges1996, Kunes2011, Boehnke2011}
requires a more general object -- the 2PCF
$X_{ijs\underline{\nu},kls'\underline{\nu'}}(\bq,\omega)$,
where $i,j,k,l$ are the spin-orbital indices, $s,s'$ are the
sublattice indices, and $\nu,\nu'$ are fermionic indices, which
represent the imaginary time evolution, e.g. Matsubara frequencies or
imaginary time. Since the fermionic indices appear only as dummy
variables in the BSE, the equation is invariant under their unitary
transformation. In the present calculations we use the Legendre
basis~\cite{Boehnke2011} for the fermionic indices. The susceptibility
$\chi_{\ldots,\ldots}(\bq,\omega)$  is obtained by the contraction of
$X_{\dots\nu,\ldots\nu'}(\bq,\omega)$  with the basis dependent structure
factor $F_\nu$~\cite{Boehnke2011}~\footnote{$F_\nu=1$ in the Matsubara
  frequency basis.}
%%%%%%%%%%%%%
\begin{equation}
    \chi_{ijs,kls'}(\bq,\omega)=F_{\underline{\nu}}F_{\underline{\nu}'}
    X_{ijs\underline{\nu},kls'\underline{\nu}'}(\bq,\omega).
\end{equation}
\\
The 2PCF $X_{ijs\nu,kls'\nu'}(\bq,\omega)$ is the solution
of the lattice BSE (\ref{eq:bsq}) using the local 2P-irreducible 
vertices $\Gamma^{s}_{ij\nu,kl\nu'}(\omega)$ and the lattice bubbles
$X^0_{ijs\nu,kls'\nu'} (\bq,\omega)$. The vertices are related to the
impurity 2PCF $\x_{ij\nu,kl\nu'}(s;\omega)$ via the impurity
BSE (\ref{eq:bsi}) for each sublattice $s$.
%Here the impurity two-particle correlation
%function  is measured in the QMC simulation.
%\lipsum[1]
\begin{widetext}
\begin{align}
\label{eq:bsq}
&X_{ijs\nu,kls'\nu'}(\bq,\omega) = X^0_{ijs\nu,kls'\nu'}(\bq,\omega)
  + X^0_{ijs\nu,\underline{mns_1\nu_1}} (\bq,\omega)
  \Gamma^{\underline{s_1}}_{\underline{mn\nu_1},\underline{pq\nu_2}}(\omega)
  X_{\underline{pqs_1\nu_2},kls'\nu'}(\bq,\omega) 
  \\
  \label{eq:bsi}
& \x_{ij\nu,kl\nu'}(s;\omega) =
  \x^0_{ij\nu,kl\nu'}(s;\omega)
  + \x^0_{ij\nu,\underline{mn\nu_1}}(s;\omega)
  \Gamma^{s}_{\underline{mn\nu_1},\underline{pq\nu_2}}(\omega)
  \x_{\underline{pq\nu_2},kl\nu'}(s;\omega).
\end{align}
\end{widetext}
% \lipsum[1]
The lattice and local bubbles
\begin{align*}
  % \label{eq:bubble}
  & X^0_{ijs\nu,kls'\nu'}(\bq,\omega)= -\delta_{\nu\nu'}
    G_{is,ks'}(\underline{\bk}\!+\!\bq,\nu\!+\!\omega)
    G_{ls',js}(\underline{\bk},\nu)
  \\
  % \label{eq:bubblei}
  & \x^0_{ij\nu,kl\nu'}(s;\omega) = -\delta_{\nu\nu'}
    G_{is,ks}(\underline{\bk},\nu\!+\!\omega)
    G_{ls,js}(\underline{\bk'},\nu).
\end{align*}
are obtained from the 1P propagator
\begin{equation*}
  G_{is,js'}(\bk,\nu) = \left[i\nu-h_{\bk} - \Sigma(\nu)
  \right]^{-1}_{is,js'}.
\end{equation*}
Note that in the 2P quantities, such as $X_{ijs\nu,kls'\nu'}(\bq,\omega)$, 
the spin-orbital indices $i$ and $j$ ($k$ and $l$) point to the same lattice site
and thus share the sublattice index $s$ ($s'$). This is because we are interested in 
correlators of local operators, i.e., products of the type $c_{\bR
  s\ldots}^{\dagger}c_{\bR s\ldots}^{\phantom\dagger}$, and because
the DMFT vertex $\Gamma^{s}_{jk\nu,kl\nu'}(\omega)$ is local. As a
result the corresponding matrices scale with the square of the number
of spin-orbitals per site $2N$, but only linearly with the number of
sites per unit cell.

%%%%%%%%%%%

We used the 1P propagators at 300 Matsubara frequencies 
and a uniform $55\times55$ ($25\times25\times25$) $k$-mesh in the 2D (3D) case
to compute the lattice and local bubbles $X_{ijs\nu,kls'\nu'}(\bq,\omega)$ and
$\x^0_{ij\nu,kl\nu'}(s;\omega)$. These are then transformed into the
Legendre representation~\cite{Boehnke2011}.
The 2P correlation function $\x_{ij\nu,kl\nu'}(s;\omega)$ is sampled
using the CT-QMC directly in the Legendre basis. 
The local 2P-irreducible vertex $\Gamma^{s}_{ij\nu,kl\nu'}(\omega)$ 
is obtained from  the impurity BSE (\ref{eq:bsi}). 
Next we solve the lattice BSE (\ref{eq:bsq})
independently for each bosonic Matsubara frequency and
 $\bq$-point. We use from 22 (for the zeroth bosonic
frequency) to 45 Legendre coefficients (for the fourteenth bosonic
frequency). A sizable reduction of the computational and storage cost
can be achieved with the procedure of
Refs.~\onlinecite{SpM,shinaoka2020}. 
 
We found that 15 bosonic Matsubara frequencies allow
for a stable analytic continuation in the low-energy region, which for
the chosen interaction strengths dominates over the high-energy
particle-hole continuum, which we do not attempt to extract. In
particular, it allows to detect the opening of spin gaps and the
suppression of otherwise divergent spectral weights caused by minor
lowering of the Hamiltonian symmetry.
Given the insufficient data statistics we have used a conservative
estimate of the uncorrelated standard deviations of 0.02 at each Matsubara
frequency point~\cite{Kaufmann2021b}. Similar calculations for weaker
interaction strength, for which the particle-hole continuum co-exists
with magnon modes, would require more Matsubara frequencies.

The 
%final output presented below are the 
spectral functions
$B^{\alpha\alpha}_{\bk}(\omega)=-\tfrac{1}{\pi}\Im\chi^{\alpha\alpha}(\bq,\omega)$ 
are our final product. Given the divergence of $B_{\bk}(\omega)$ at the ordering
wave vector, we introduce an effective cutoff by plotting
$\tilde{B}_{\bk}(\omega)=\tfrac{B_{\bk}(\omega)}{C+B_{\bk}(\omega)}$ instead of
 $B_{\bk}(\omega)$ itself. To represent the amplitude we also plot the
 spectral weight $\Omega_{\bk}=\int_0^{0.5}B_{\bk}(\omega)d\omega$.

\begin{figure}
  \begin{minipage}[b]{0.005\columnwidth}
    \hfill\vspace{\textwidth}
  \end{minipage}
  \begin{minipage}[b]{0.654\columnwidth}
    \includegraphics[width=\textwidth]{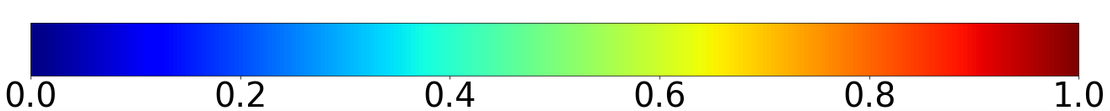}
  \end{minipage}
   \begin{minipage}[b]{0.235\columnwidth}
     \hfill\vspace{\textwidth}
   \end{minipage}
  \begin{minipage}[b]{0.742\columnwidth}
    \includegraphics[width=\textwidth]{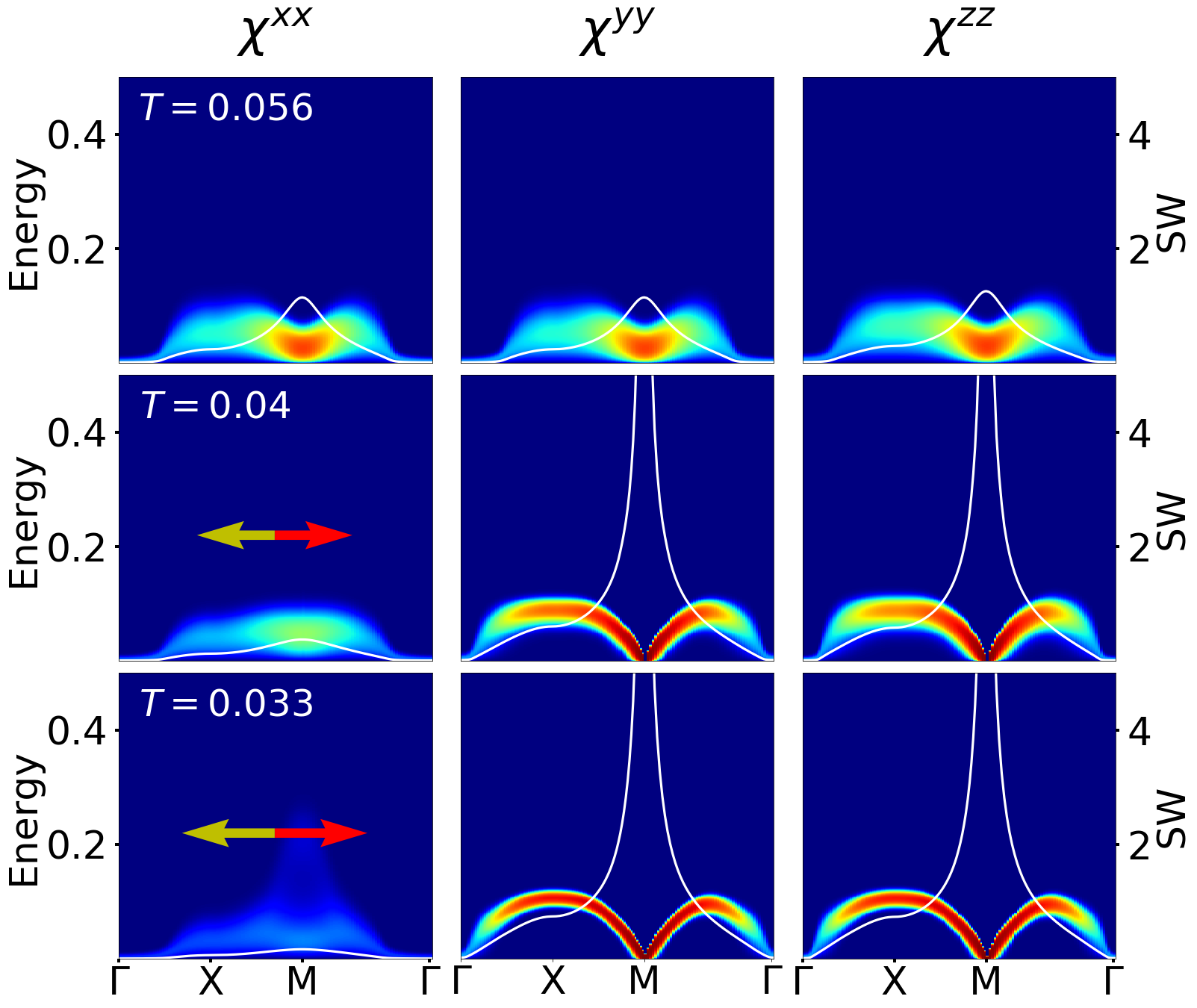}
    \vspace{-0.13cm}
  \end{minipage}
  \begin{minipage}[b]{0.240\columnwidth}  
    \includegraphics[width=\textwidth]{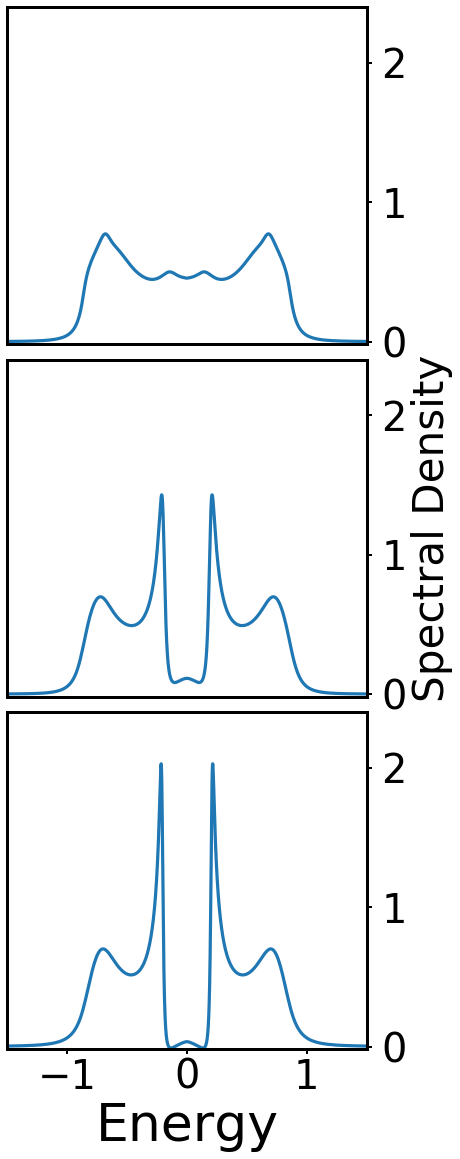}
  \end{minipage}
  \caption{The AFM ordering in the single-band Hubbard model. Left:
    Imaginary part of the dynamical susceptibility $\chi(\bq, \omega)$
    ($\tilde{B}_{\bk}(\omega)$ with $C=5.5$ is plotted)
    along the $\Gamma(0,0)$--$X(\pi,\pi)$--$M(2\pi,0)$ path in 
    the extended Brillouin zone for the 2-site unit cell
    plotted as color plot for 
    various temperatures. The top row corresponds to the normal state
    at $T\approx 1.12 T_c$. The lower rows are obtained in the AFM
    state with $\bN$ along the $x$-axis. The white line (right axis) shows the
    integrated spectral weight $\Omega_{\bk}$. Right: The corresponding 1P spectral
    functions.}
    \label{fig:single}
\end{figure}

\begin{figure}
\centering
   \includegraphics[width=0.8\columnwidth]{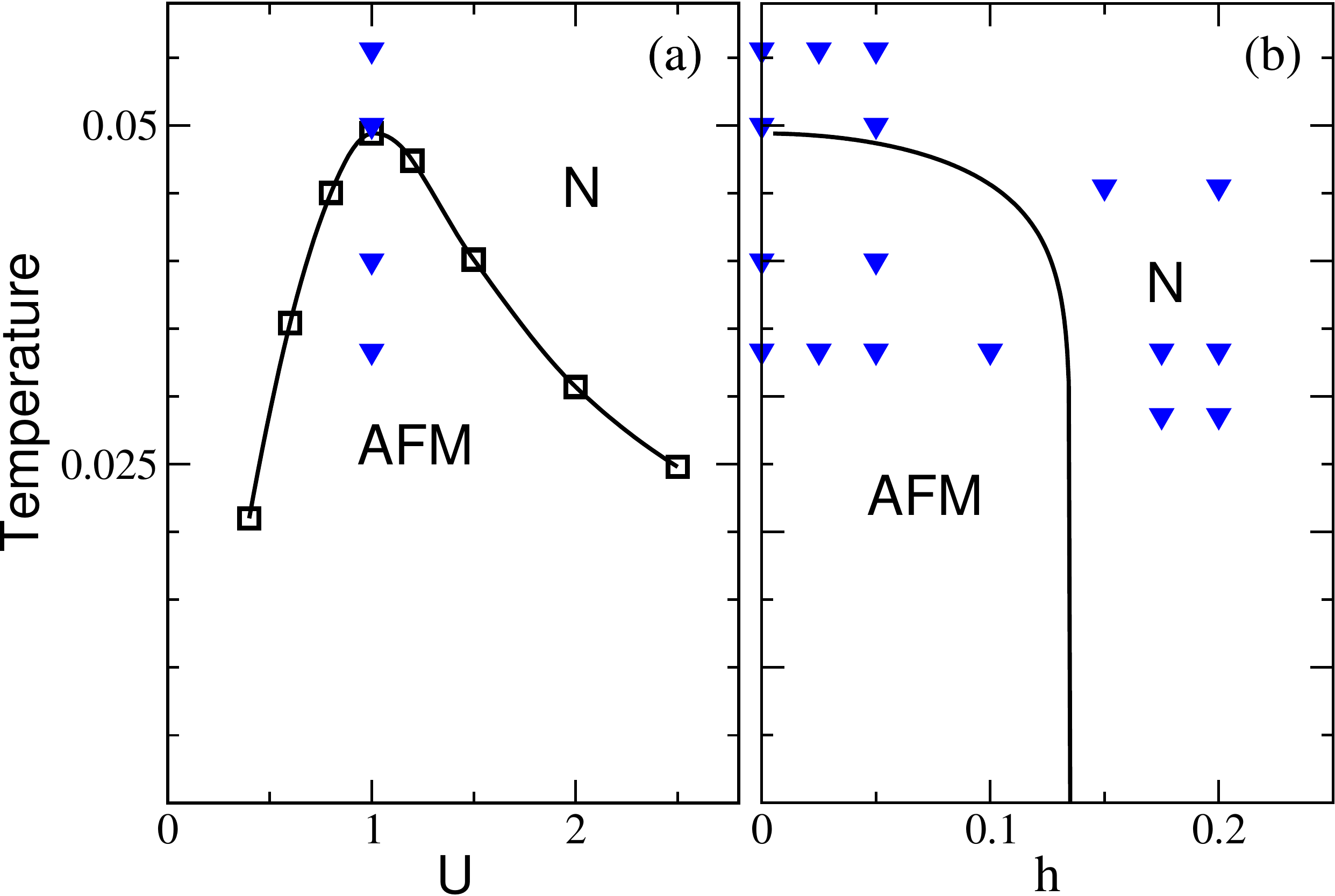}
    \caption{(a) The phase boundary between AFM and normal (N) phase adapted from
      Ref.~\onlinecite{Kunes2011}. (b) The schematic
      phase diagram in the $T-h$ plane ($U=1$). Symbols mark the points of
      actual calculations.}
    \label{fig:phase_diag}
\end{figure}

\begin{figure}
  \begin{minipage}[b]{0.001\columnwidth}
    \hfill\vspace{\textwidth}
  \end{minipage}
  \begin{minipage}[b]{0.626\columnwidth}
    \includegraphics[width=\textwidth]{cbar1.png}
  \end{minipage}
  \begin{minipage}[b]{0.236\columnwidth}
    \hfill\vspace{\textwidth}
  \end{minipage}
  \begin{minipage}[b]{0.712\columnwidth}
    \includegraphics[width=\textwidth]{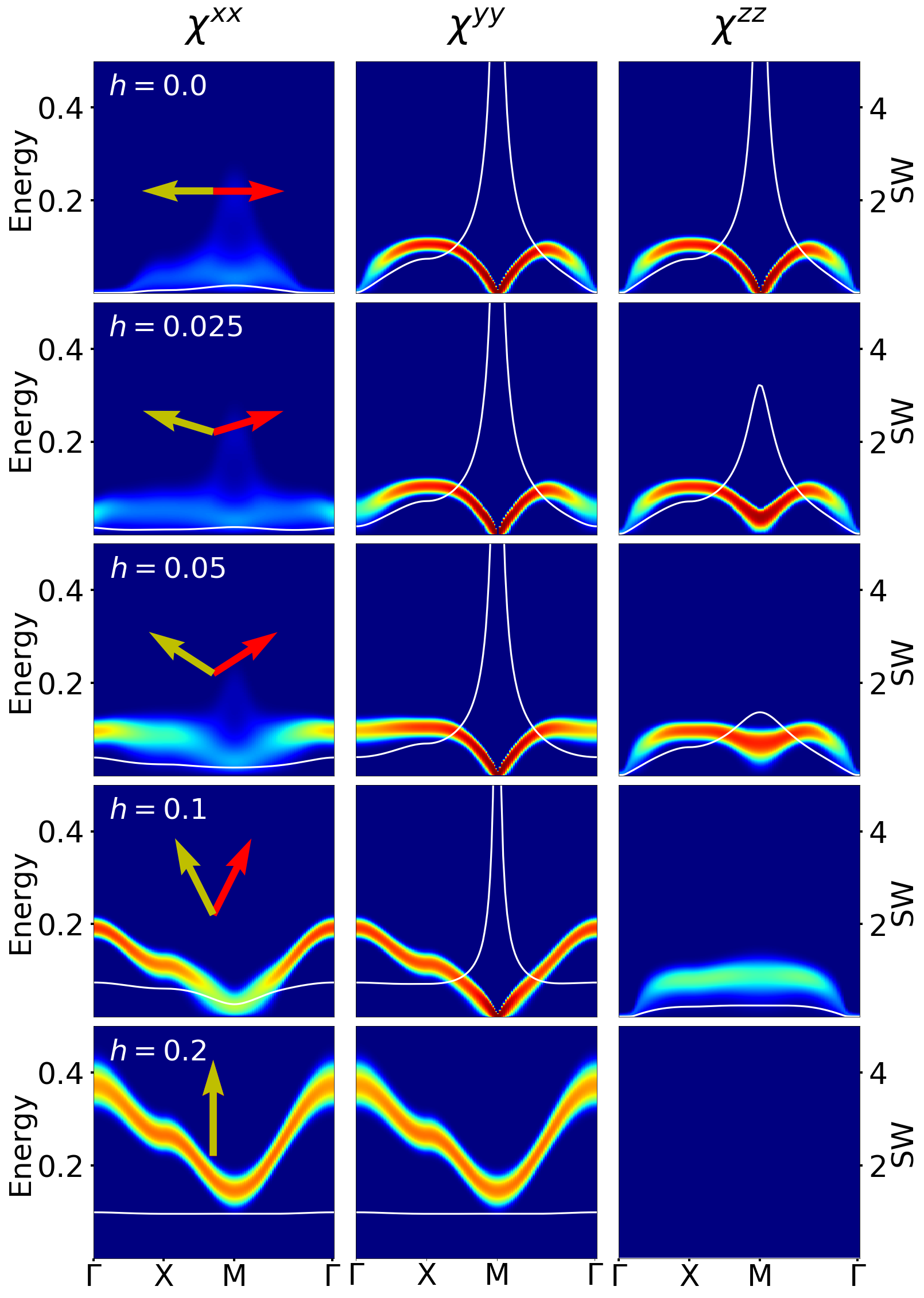}
    \vspace{-0.13cm}
  \end{minipage}
  \begin{minipage}[b]{0.245\columnwidth}
    \includegraphics[width=\textwidth]{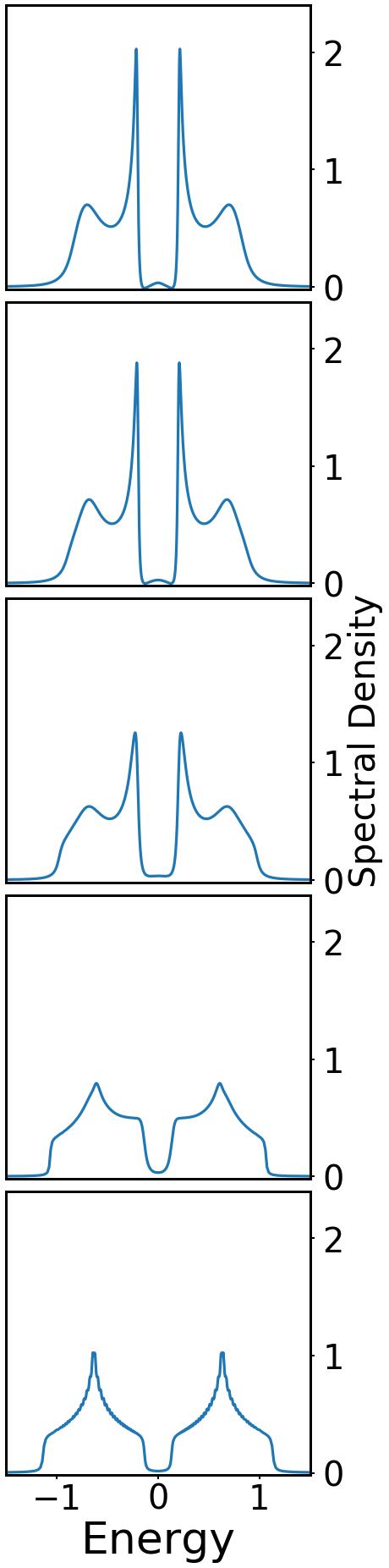}
  \end{minipage}
  \caption{
    The same as in Fig.~\ref{fig:single} 
    for various amplitudes of the external field $\bh$
    (along the $z$-axis) at $T=1/30$.
    The top row corresponds to the AFM state without field, the bottom
    row to a fully polarized normal state. The arrows show the
    size and tilt of the sublattice spin polarization.
    The N\'eel vector 
    $\bN$ points along the $x$-axis.}
  \label{fig:h-field}
\end{figure}

\section{Results and Discussion}

\subsection{S=1/2 and magnetic field}
First, we discuss the antiferromagnetism of the single orbital 2D model.
The local term in Eq.~\ref{eq:model} adopts the form
\begin{equation}
\label{eq:o1}
  H_i = U n_{i\uparrow}n_{i\downarrow}+ h\left(n_{i\uparrow}
    - n_{i\downarrow} \right).
\end{equation}
At $U$=1 the model is close to the maximal transition temperature
between the weak-coupling RPA and strong-coupling Heisenberg
regimes~\cite{Kunes2011,Rohringer2016,Schafer2021}. In Fig.~\ref{fig:single} we show the evolution of the
electron spectral density and the dynamical spin susceptibility across
the AFM transition along with the local 1P spectral densities. 
The location of the studied temperatures
in the phase diagram is shown in Fig.~\ref{fig:phase_diag}a. The
direction of the staggered magnetization,
$\expval{\bS_\bR}=(-1)^{|\bR|}\bN$ is chosen along the
$x$-axis. The elements $\chi^{yy}$ and $\chi^{zz}$ reflect the two
linear Goldstone modes~\cite{Watanabe2012} arising from breaking of
$SU(2)$ symmetry in an antiferromagnet. 
Note, that their spectral weight diverges at the ordering wave vector ($M$-point). 
While the Hamiltonian
(\ref{eq:model}) for $h=0$ is isotropic in spin space, the
numerical treatment of $\chi^{yy}$ is kept independent of that of
$\chi^{zz}$. Nevertheless, the results reflect the
symmetry quite accurately.

\begin{figure}
  \centering
  \includegraphics[width=0.9\columnwidth]{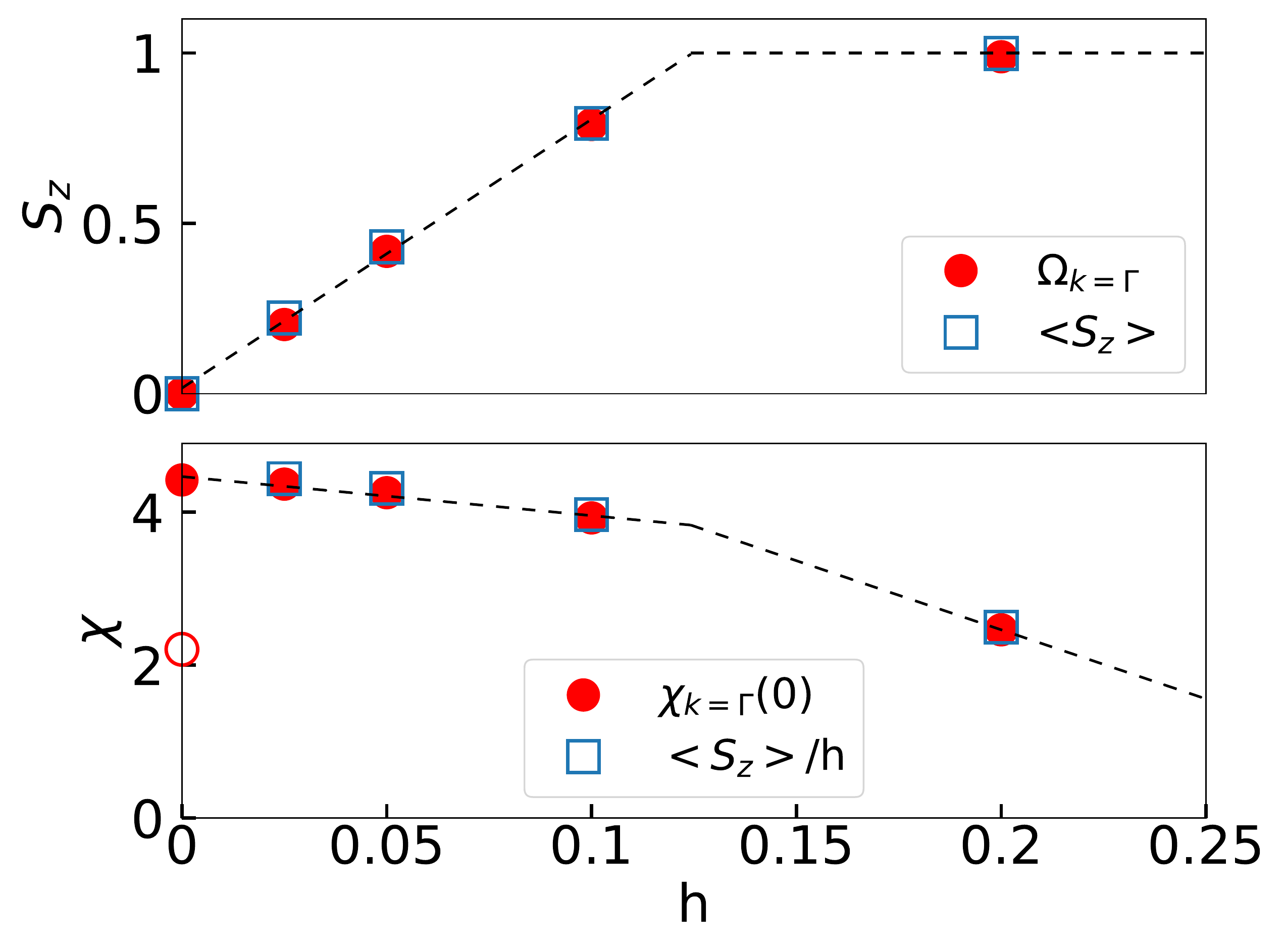}
  \caption{(top) The spectral weight of the transverse ($xx$ or $yy$) magnons
    of Fig.~\ref{fig:h-field} compared to the induced moment $\langle S_z\rangle$.
    (bottom) The static uniform transverse susceptibility $\chi^{xx}(0,0)=\chi^{yy}(0,0)$ compared
    to its exact value $\tfrac{\langle S_z\rangle}{h}$. The dotted line in the top 
    panel is piecewise: a quadratic fit to the data and a constant (reflecting the
    spin saturation). In the lower panel the same curve is divided by $h$.
    Note that at $h=0$ the equivalence of $\chi^{xx}$ and $\chi^{yy}$ 
    is broken. The empty circle marks the value of $\chi^{xx}(0,0)$, while
    the full circle that of $\chi^{yy}(0,0)$.}
  \label{fig:swh}
\end{figure}

Application of a uniform magnetic field is known to polarize an isotropic
antiferromagnet perpendicularly to the field, $\bN\perp\bh$, while the
spin density acquires a uniform component parallel to the field:
$\expval{\bS_\bR}=\bM+(-1)^{|\bR|}\bN$, with $\bM \parallel \bh$. The symmetry of the Hamiltonian
is reduced from $SU(2)$ to $U(1)$~\footnote{We mention only the spin
  symmetries, which are broken or varied in our models and ignore the
  ones, which are not changed such the global $U(1)$ due to charge
  conservation.}. Its breaking leads to a single linear Goldstone mode
($\chi^{yy}$) with polarization perpendicular to both $\bN$ and $\bM$,
which corresponds to spin rotation in the $xy$-plane. Indeed, the
($\chi^{zz}$) mode polarized along the field direction is gapped, as
shown in Fig.~\ref{fig:h-field}. At the same time its spectral weight
at the $M$-point becomes finite.
%, while the magnetic moments are tilted in the field direction. 
$\chi^{yy}$ acquires a finite spectral
weight at the $\Gamma$-point at energy equal to $2h$~\cite{krien2018}. With increasing
$h$ and progressing tilt of the local moments, $\chi^{zz}$ loses
its spectral weight while the weight of $\chi^{xx}$ grows from the
$\Gamma$-point outwards. These observations agree with zero temperature
exact diagonalization of Ref.~\onlinecite{Luschner2009}, although
multi-magnon satellites are as expected absent in the DMFT
spectra. Eventually, the AFM order is lost, as shown in
Fig.~\ref{fig:phase_diag}b, the system becomes fully spin-polarized
and recovers the $U(1)$ symmetry of its Hamiltonian. The perpendicular
susceptibilities ($\chi^{xx}$ and $\chi^{yy}$) describe gapped magnons
(perhaps better called magnetic excitons - a spin flip propagating in
a spin-polarized background). The field-driven transition can be
viewed as the Bose-Einstein condensation of these magnons, which takes
place when the magnon gap is closed. The direction of $\bN$ within the
$xy$-plane plays the role of the condensate phase.

The uniform susceptibilities (at the $\Gamma$-point) 
%in Fig.~\ref{fig:h-field} 
reflect the conservation of the total spin moment along the field
direction ($S^z$), which leads to (i)
${\chi^{zz}(0,\omega) = \beta \expval{(S^z)^2} \delta(\omega)}$ having
no dynamics and therefore vanishing imaginary part, (ii)
${\chi^{xx}(0,\omega)=\chi^{yy}(0,\omega)}$. For a simple proof see
the Appendix~\ref{sec:appA}. Our empirical observations showed that
the numerical noise in $\chi^{\alpha\alpha}(\bq,\omega_n)$ is essentially
independent of $\bq$. This is consistent with the fact that such noise
mostly originates from the QMC calculation of the local 2PCF,
which is used in the determination of the local irreducible vertex. As
a consequence, one may subtract
$\chi^{\alpha\alpha}(\bq,\omega_n) - \chi^{\alpha\alpha}(0,\omega_n)$
for $\omega_n>0$ in order to reduce the noise in cases where (i)
holds.

The uniform susceptibility in external field offers further simple
consistency tests. First, its static part is exactly
$\chi^{xx}(0,0)=\tfrac{\langle S_z\rangle}{h}$, reflecting the fact
that application of an infinitesimal transverse field simply rotates
the net moment in the new field direction. Second, the spectral weight
is equal to the net moment $\Omega^{xx}_0=\langle S_z\rangle$ (see
Eq. D.4 of Ref.~\onlinecite{krien2018}). Fig.~\ref{fig:swh} shows that
our numerical results respcect these properties with great accuracy
across all field values.

\subsection{S=1 and single-ion anisotropy}
\begin{figure}
  %\begin{minipage}[b]{0.02\columnwidth}
  %  \includegraphics[width=\textwidth]{}
  %\end{minipage}
  \begin{minipage}[b]{0.82\columnwidth}
    \includegraphics[width=\textwidth]{cbar1.png}
  \end{minipage}
  %\begin{minipage}[b]{0.20\columnwidth}
  %  \includegraphics[width=\textwidth]{}
  %\end{minipage}
  \begin{minipage}[b]{1.0\columnwidth}
    \includegraphics[width=0.95\linewidth]{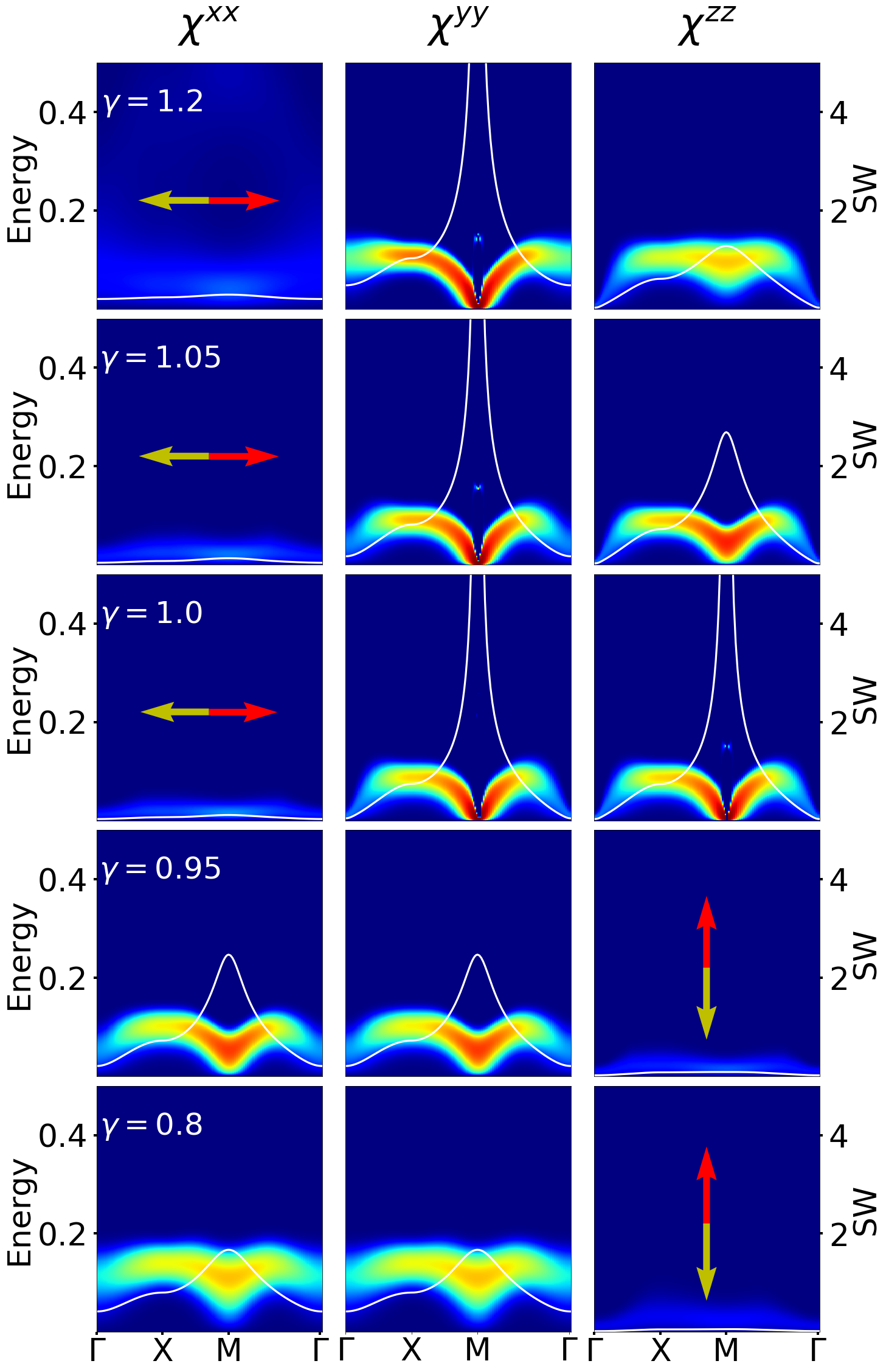}
    \caption{
      Imaginary part of the dynamical susceptibility $\chi(\bq,\omega)$ 
      ($\tilde{B}_{\bk}(\omega)$ with $C=8.5$ is plotted)
      at $T=1/30$ in the AFM state of two-orbital Hubbard model for
      various single-ion anisotropy. The rows 1-2 correspond to easy
      ($xy$) plane, row 3 to $SU(2)$ symmetry, and rows 4-5 to easy
      ($z$) axis anisotropy. The white line (right axis) shows the
      the integrated spectral weight $\Omega_{\bk}$.}
    \label{fig:2-orb-aniso} 
  \end{minipage}
\end{figure}

Next, we investigate the effect of single-ion anisotropy on the magnon
dispersion. To this end we study a two-orbital model (at half filling
$n=2$)
\begin{equation}
\label{eq:o2}
\begin{split}
  H_i =& U \sum_{l=1,2}n_{il\uparrow}n_{il\downarrow}+
  U'\sum_{\sigma,\sigma'}n_{i1\sigma}n_{i2\sigma'}\\
  -&J \sum_{\sigma} \left(
    n_{i1\sigma}n_{i2\sigma} 
    +\gamma
    c_{i1\sigma}^{\dagger}c_{i1\bar{\sigma}}^{\phantom\dagger}
    c_{i2\bar{\sigma}}^{\dagger}c_{i2\sigma}^{\phantom\dagger} \right)
\end{split}
\end{equation}
with $U=1$, $J=0.25$ and $U^{\prime} = 0.5$
\footnote{{For the sake of simplicity we have omitted the pair-hopping
    term $c_{i1\uparrow}^{\dagger} c_{i1\downarrow}^{\dagger}
    c_{i2\downarrow} c_{i2\uparrow}+H.c.$  This term does not affect
    the broken spin symmetry and is expected to play a minor role when
    the local high-spin state dominates as in the AFM phases studied
    here.}}
The single ion anisotropy is introduced by unequal weights of the Ising
and spin-flip terms in the interaction Hamiltonian. This way 
the $SU(2)$ symmetry ($\gamma=1$) is reduced to a $Z_2\times U(1)$
for $\gamma\neq 1$. The residual symmetry of the AFM state 
depends on $\gamma$.  For $\gamma>1$ the atomic ground state
corresponds to $\ket{S,S_z}=\ket{1,0}$, i.e., a state with no spin
dipole moment. However, for moderate deviations $\gamma\gtrsim 1$ the
inter-site exchange, which favors (dipole) magnetic order, enforces AFM
order within the $xy$-plane. The in-plane order breaks the $U(1)$ 
symmetry and thus one linear Goldstone mode is expected.  For
$\gamma<1$ the atomic ground state corresponds to
$\ket{S,S_z}=\ket{1,\pm 1}$, i.e., an Ising ground state. The
inter-site exchange leads to the formation of AFM order with moments along
the $z$-axis. The residual symmetry of the ordered state is $U(1)$ and
only the discrete $Z_2$ symmetry is broken at the transition. Therefore no
gapless Goldstone mode is expected. The numerical results are presented
in Fig.~\ref{fig:2-orb-aniso}. In the $SU(2)$ symmetric case
we have numerically tested the stability of the AFM state with arbitrary
orientation of the N\'eel vector $\bN$. We observe two linear Goldstone modes in
the response to a field perpendicular to the N\'eel vector, i.e.,
$\chi^{yy}$ and $\chi^{zz}$ for $\bN\parallel x$. For
$\gamma\neq 1$ the system self-consistently picks the
expected N\'eel vector. For $\gamma>1$ with the in-plane order, we
choose $\bN\parallel x$. As in the $SU(2)$
symmetric case we find minor longitudinal response %($\chi^{xx}$ in this case) 
and progressive gapping of the out-of-plane $\chi^{zz}$
component with increasing $\gamma$. For $\gamma<1$ the system picks $\bN\parallel z$. 
In this case, the residual $U(1)$ symmetry is reflected
in the equivalence of $\chi^{xx}=\chi^{yy}$, both of which are
progressively gapped when lowering $\gamma<1$.

The presence of a spin gap is known to stabilize the long-range order in 2D. On the other
hand, linear gapless magnon mode is detrimental to the long-range
order at any finite temperature~\cite{Mermin1966} if the thermal
population of this mode is properly taken into account.
This is not the case of the DMFT treatment, in which the low-energy
long-range spin fluctuations do not feed back to the calculation of
the 1P and 2P vertices. Nevertheless, the present results 
show that DMFT accurately captures the behavior of the spin
gap. Therefore it provides a useful reference and a starting point for
more sophisticated approaches such as D$\Gamma$A~\cite{Toschi2007,Kaufmann2021} or dual
fermions~\cite{Rubtsov2008,Rohringer2018}. The model with tunable $\gamma$ not only hosts
both states with gapped and gappless magnons, but allows a continuous
tuning between them. Investigation of $T_c(\gamma)$ dependence may
thus provide a useful test of the above methods as well as a
quantitative measure of their accuracy.

\subsection{S=3/2 and spin-orbit coupling}
\begin{figure}
   \includegraphics[width=0.95\linewidth]{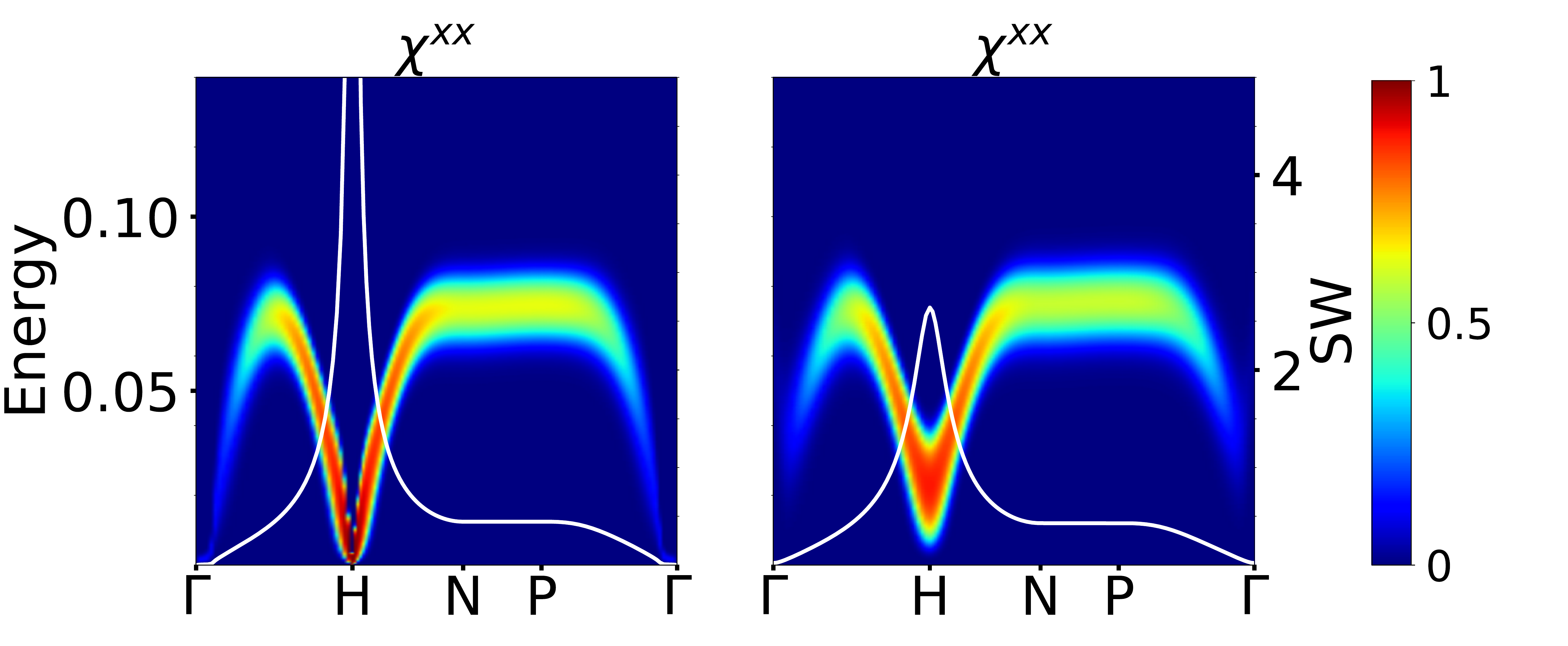}
  \caption{Imaginary part of the transverse dynamical susceptibility
    $\chi(\bq,\omega)$  ($\tilde{B}_{\bk}(\omega)$ with $C=20$ is plotted)
    at $T=1/40$ in the AFM state of 3-orbital
    Hubbard model ($\bN \parallel z$-axis) with (right) and without
    (left) the spin-orbit coupling.
    The data are plotted along the
    $\Gamma(0,0,0)$--$H(2\pi, 0, 0)$--$N(\pi, \pi, 0)$--$P(\pi, \pi, \pi)$
    path in the extended Brillouin zone for the 2-site unit cell.
    The white line (right axis) shows the
    the integrated spectral weight $\Omega_{\bk}$.
    }
  \label{fig:susc_cubic}
\end{figure}

Finally, we study the more realistic 3D case of a three orbital model on a
bcc cubic lattice and introduce spin-orbit coupling as the source
of spin anisotropy
\begin{equation}
\label{eq:o3}
\begin{split}
  H_i =&
  \sum_{l,l'}\sum_{\sigma,\sigma'}h^{\text{soc}}_{l\sigma,l'\sigma'}
  c_{il\sigma}^{\dagger} c_{il'\sigma'}^{\phantom\dagger}\\
  +&U \sum_{l} n_{il\uparrow} n_{il\downarrow}
  + U'\sum_{l>l'} \sum_{\sigma,\sigma'}
  n_{il\sigma} n_{il'\sigma'}\\
  -&J \sum_{l>l'} \sum_{\sigma} \left(n_{il\sigma}n_{il'\sigma} 
    + c_{il\sigma}^{\dagger} c_{il\bar{\sigma}}^{\phantom\dagger}
    c_{il'\bar{\sigma}}^{\dagger} c_{il'\sigma}^{\phantom\dagger}
      \right),
\end{split}
\end{equation}
where $U=2$, $J=0.5$ and $U^{\prime}=1$
The form of $h^{\text{soc}}_{l\sigma,l'\sigma'}$ is given
in Appendix~\ref{sec:appB}. We point out that the present CT-QMC
calculation may become impossible at low temperatures due to the sign problem
associated with the
spin-orbit coupling~\cite{Kim2020}. Nevertheless, in the half-filled
case the effect of spin-orbit coupling is rather 
moderate~\cite{Kim2020,Ahn2017} and we are able to reach the AFM phase
without problems. The N\'eel vector picks an orientation along a cubic axis ($\bN\parallel z$).
In Fig.~\ref{fig:susc_cubic} we show the calculated transverse
susceptibilities. Without spin-orbit coupling we observe a linear
Goldstone mode. The spin-orbit coupling leads to the opening of a finite
spin gap. In the 3D case, the AFM order is physically realistic. The DMFT 
transition temperature $T_c$ provides a realistic estimate, which
misses corrections due to long-range spin fluctuations, while in 2D
the corrections dominate.
% \begin{figure}
%   \includegraphics[width=0.45\columnwidth]{cubic_nosoc.pdf}
%   \includegraphics[width=0.45\columnwidth]{cubic_soc.pdf}
%   \caption{Imaginary part of the transverse dynamical susceptibility
%     $\chi(\bq,\omega)$ at $T=1/40$ in the AFM state of 3-orbital
%     Hubbard model
%     ($\bN\parallel z$-axis) with (right) and without (left) the 
%     spin-orbit coupling.}
%   \label{fig:susc_cubic}
% % \end{figure}

\begin{comment}
\subsection{Discussion}
In the present multi-orbital calculations we have omitted so called
pair-hopping term in the interaction Hamiltonian. This does not affect
the studied symmetry breaking and has overall minor effect in
half-filled high-spin systems.

As usual with imaginary time methods, analytic continuation poses a
severe problem especially at high energies. Despite using only 15
bosonic Matsubara frequencies, we have avoided it by focusing on the
low-energy spin response, which dominates for chosen interaction
strengths. Similar calculations for weaker interaction, for which the
particle-hole continuum co-exists with magnon modes, would require
more Matsubara frequencies.  We have shown that even minor lowering of
the Hamiltonian symmetry results in opening of a spin gap and
suppression of otherwise divergent spectral weight.
Uniform susceptibility in finite external field provides a
particularly useful benchmark for the linear response calculations
since it can be calculated directly from occupation numbers.
\end{comment}

\section{Conclusions}
We have presented DMFT calculations of the AFM phase of the half-filled
Hubbard model with one, two and three orbitals in the
intermediate coupling regime in two and three dimensions. We find
that the expected behavior of magnons in response to external magnetic
field or single-ion anisotropy is well captured. While the 3D
description of AFM ordering is physically relevant, the ordering
behavior in 2D is not correct since DMFT is not sensitive to
dimensionality and violates the Mermin-Wagner theorem, which prohibits
spontaneous breaking of continuous symmetry at finite
temperature. Nevertheless, the fact that DMFT magnons properly
describe the opening of the spin gap suggests that DMFT is a good
starting point of theories, which properly include the long-range spin
fluctuations. The two-orbital model with variable single-ion
anisotropy provides an ideal test case for such theories as it allows
one to continuously tune between the breaking of a discrete and of a
continuous symmetry. On the computational level we have shown that
3-orbital calculations, which cover for example models of
ruthenates or iridates, are numerically feasible.

\begin{acknowledgements}
The authors thank A.~Kauch, F.~Krien, J.~Kaufmann, A.~Toschi and
M.~Wallerberger for comments and critical reading of the
manuscript. This work was supported by the European Research Council
(ERC) under the European Union's Horizon 2020 research and innovation
programme (grant agreement No.~646807-EXMAG). D.G was supported by the
Czech Science Foundation (GA\v{C}R) under Project No.~GA19-16937S.
The authors acknowledge support by the Czech Ministry of Education,
Youth and Sports from the Large Infrastructures for Research,
Experimental Development and Innovations project \textquotedblleft
IT4Innovations National Supercomputing Center –
LM2015070\textquotedblright. Part of the calculations were performed
at the Vienna Scientific Cluster.
\end{acknowledgements}

\begin{appendix}
\section{}
\label{sec:appA}
The susceptibilities at the $\Gamma$-point correspond to correlators
of total spin momenta $S^\alpha=\sum_\bR S^\alpha_\bR$. The
Hamiltonian (\ref{eq:o1}) commutes with $S^z$, $[S^z,H]=0$. First, we show
that the $zz$ correlator does not depend on the imaginary time $\tau$
\begin{equation*}
    \expval{ S^z(\tau)S^z} \equiv 
    \expval {e^{\tau H} S^z e^{-\tau H} S^z }=\expval{ (S^z)^2}.
    \end{equation*}
This implies that only $\chi^{zz}(0,\nu_n=0)$ is finite and equal to
$\beta \expval{(S^z)^2}$.

To prove the equality of $\chi^{xx}$ and $\chi^{yy}$ we write the
corresponding spin-spin correlation functions with the help of ladder
operators ${S^x=S^-+S^+}$ and ${S^y=i(S^--S^+)}$:
\begin{equation*}
\begin{split}
    \expval{ S^x(\tau)S^x}=&
    \expval{ e^{\tau H}S^+e^{-\tau H}S^-}+
    \expval{ e^{\tau H}S^-e^{-\tau H}S^+}\\
    &+
    \expval{ e^{\tau H}S^-e^{-\tau H}S^-}+
    \expval{ e^{\tau H}S^+e^{-\tau H}S^+}
  \end{split}
\end{equation*}
\begin{equation*}
  \begin{split}
    \expval{ S^y(\tau)S^y}=&
    \expval{ e^{\tau H}S^+e^{-\tau H}S^-}+
    \expval{ e^{\tau H}S^-e^{-\tau H}S^+}\\
    &-
    \expval{ e^{\tau H}S^-e^{-\tau H}S^-}-
    \expval{ e^{\tau H}S^+e^{-\tau H}S^+}.
  \end{split}
\end{equation*}
Since $H$ commutes with $S^z$ only the $+-$ and $-+$ contributions are
non-zero, while the $++$ and $--$ contributions are equal to zero,
which implies the $\Gamma$-point equality
$\chi^{xx}(0,\omega)=\chi^{yy}(0,\omega)$.

\section{}
\label{sec:appB}
The spin-orbit coupling mimics that in the $t_{2g}$ subspace of
atomic $d$-shell
%is taken from Eq. (7) of Ref.
~\cite{du_metal-insulator_2013}
commonly found in real materials (with cubic site-symmetry).
In the on-site basis  
%$(d_{yz\uparrow}, d_{yz\downarrow},
%d_{zx\uparrow}, d_{zx\downarrow}, d_{xy\uparrow}, d_{xy\downarrow})$
${\{1\!\!\uparrow, 1\!\!\downarrow,
2\!\!\uparrow, 2\!\!\downarrow, 3\!\!\uparrow, 3\!\!\downarrow\}}$
it reads
\begin{equation*}
  h^{\text{SOC}} = -\dfrac{\zeta}{2}
  \begin{pmatrix} 
    0 & 0 & -i & 0 & 0 & 1 \\
    0 & 0 & 0 & i & -1 & 0 \\
    i & 0 & 0 & 0 & 0 & -i \\
    0 & -i & 0 & 0 & -i & 0 \\
    0 & -1 & 0 & i & 0 & 0 \\
    1 & 0 & i & 0 & 0 & 0 \\
  \end{pmatrix}
  .
\end{equation*}
The calculations presented in this work use 
%a weak value of the spin-orbit coupling 
$\zeta=0.1$.

\end{appendix}

\bibliography{supersolid}

\end{document}